\documentstyle[epsf,epsfig,11pt]{article}
\begin{document}

\renewcommand{\arraystretch} {1.3}

\def\NPB#1#2#3{{\rm Nucl.~Phys.} {\bf{B#1}} (#2) #3}
\def\PLB#1#2#3{{\rm Phys.~Lett.} {\bf{B#1}} (#2) #3}
\def\PRD#1#2#3{{\rm Phys.~Rev.} {\bf{D#1}} (#2) #3}
\def\PRL#1#2#3{{\rm Phys.~Rev.~Lett.} {\bf{#1}} (#2) #3}
\def\EPJ#1#2#3{{\rm Eur.~Phys.~J.} {\bf C#1} (#2) #3}
\def\CPC#1#2#3{{\rm Comp.~Phys.~Comm.} {\bf#1} (#2) #3}
\def\JHEP#1#2#3{{\rm JHEP} { \bf{#1}} (#2) #3}
\newcommand{\etal}{{\it et al.}}

\def\lt{\raisebox{0.2ex}{$<$}}
\def\gt{\raisebox{0.2ex}{$>$}}

\def\GeV{\ifmmode {\mathrm{Ge\kern -0.1em V}}\else
                   \textrm{Ge\kern -0.1em V}\fi}%
\def\TeV{\ifmmode {\mathrm{TeV}}\else
                   \textrm{TeV}\fi}%
\def\antibar#1{#1\bar{#1}}%
\def\ee{\ifmmode {e^+e^-}\else
                 {$e^+e^-$}\fi}%
\def\pp{\ifmmode {pp}\else
                 {$pp$}\fi}%
\def\ZZ{\ifmmode {ZZ}\else
                 {$ZZ$}\fi}%
\def\ZZdecay{\ifmmode {ZZ \rightarrow \ell \ell \nu \nu}\else
                      {$ZZ \rightarrow \ell \ell \nu \nu$}\fi}%
\def\ZG{\ifmmode {Z + G}\else
                 {$Z + G$}\fi}%
\def\ZU{\ifmmode {Z + U}\else
                 {$Z + U$}\fi}%
\def\ZGU{\ifmmode {Z + G/U}\else
                  {$Z + G/U$}\fi}%


\begin{titlepage}

\pagenumbering{arabic}

\begin{flushright}
       MAN/HEP/2008/28 \\
       September 2008 \\
\end{flushright}

\vspace*{1cm}

\begin{center}
\boldmath
{\Large \bf Simulation of Z plus Graviton/Unparticle Production \\ at the LHC } \\
\unboldmath
\end{center}
\vspace*{2.0cm}
\large {\bf Stefan Ask } \\
{\it School of Physics \& Astronomy, University of Manchester,\\ 
Manchester M13 9PL, UK.} \\
{\tt E-mail: Stefan.Ask@manchester.ac.uk }

\noindent \rule{\textwidth}{0.5ex}

\vspace*{2.0cm}

\begin{abstract}
\noindent
Theories with extra dimensions have gained much interest in recent years as 
candidates for a possible extension of the SM. The observation of large extra 
dimensions through real graviton emission is one of the most popular related 
new phenomena. The main experimental signatures from graviton emission are 
production of single jet and single photon events, which have been studied 
in great detail. This work describes the implementation of graviton production 
together with either a $Z$ or a photon in Pythia 8. The potential of using $Z$ 
plus graviton production at the LHC as a complementary channel is also studied. 
For completeness, this work also includes the more recently proposed scenario 
of unparticle emission, since the effective theory of unparticles to some 
extent represents a generalization of the large extra dimension model.

\end{abstract}

\vspace{2.cm}

\vspace{\fill}
\end{titlepage}

\pagebreak

\setcounter{page}{1}


\section{Introduction}

The possibility of observing gravity in extra dimensions (ED) at the \TeV\ scale 
was proposed a few years ago and has since gained a large interest. One of 
the first proposals was the so-called ADD scenario \cite{bib:ADD} which suggests 
that gravity alone would propagate in large extra dimensions (LED). This would 
explain the weakness of gravity experienced by the SM fields, since they only 
would have access to the normal 4 dimensional space-time. This scenario has been 
studied in detail \cite{bib:LED1,bib:LED2} and dedicated searches have been 
performed at different experiments \cite{bib:LEDEXP1, bib:LEDEXP2, bib:LEDEXP3}. 
The large extra dimensions in the ADD scenario have a flat space topology and, for 
this reason, astrophysical observations imply stringent constraints in the case 
of a small number of extra dimensions. It has, however, later been shown that 
these constraints would be evaded if space-time is slightly warped, as in the 
popular RS models \cite{bib:RS}. In this case the graviton ($G$) would acquire 
a small effective mass, which makes the model \cite{bib:LED3} insensitive to 
astrophysical constraints from low-energy processes\footnote{The high-energy 
phenomena relevant at collider experiments are, however, unchanged.}. Then even 
one large extra dimension could have escaped the present experimental measurements 
\cite{bib:LEDEXP4}.

Recently a so-called unparticle ($U$) model has also gained much attention 
\cite{bib:UNP1,bib:UNP2}. This relates to phenomena from a scale invariant sector 
which is coupled to the SM by a connector sector with a high mass scale. This 
scenario is normally considered to be less well motivated than extra dimensional 
gravity, which could solve the so-called hierarchy problem and is also motivated 
by string theory. It could, however, imply unusual experimental signatures at the 
LHC which should not be missed. In addition, the unparticle model is, from a 
phenomenological point of view, a generalization of the large extra dimension case 
and therefore both cases can to some extent be covered at the same time.

The main experimental signatures from both gravity in large extra dimensions as 
well as the unparticle model would be an excess of single jet and single photon 
events. In this work we describe the implementation of $Z/\gamma + G/U$ production 
as a so-called semi-internal process to Pythia 8 \cite{bib:Pythia8}. Furthermore 
\ZGU\ production at the LHC is studied, since this process could provide a complement 
in order to constrain the model if a signal would be observed in the main channels. 

The next section gives a reminder of the relevant model parameters and, in order 
to simplify comparison between different papers, the most common conventions are 
summarized. This is followed by the implementation of the process and the validation 
tests that have been performed. Finally a study at generator level of the $\pp \rightarrow \ZGU$ 
cross section predicted at the LHC is presented, including a relative comparison 
with the similar SM process $\pp \rightarrow \ZZdecay$.

\section{The Models and Parameters}
%

The extra dimension model considered in this work is the same as that described in 
\cite{bib:LED1}. Here all SM fields are confined to a 4 dimensional brane present 
in a larger dimensional space where only gravity can propagate. It is assumed 
that the brane is rigid so that effects from fields related to the brane dynamics 
can be neglected. The extra dimensions would be compactified with a radius $R$. 
In the case of real graviton production at the LHC the momentum component of the 
graviton in the extra dimensions would be observed on the brane as a mass. The 
finite size of the ED implies a discrete series of allowed mass modes, i.e. the 
so-called Kaluza-Klein (KK) tower. The graviton is coupled to the SM fields through 
the energy-momentum tensor as,
\begin{equation}
\Delta {\cal L} = -(8\pi G_N)^{\frac{1}{2}}G^{(i)}_{\mu\nu}T^{\mu\nu}
\end{equation}
where $G_N$ is the Newton constant and $i$ is the KK mode index. For the graviton 
production cross section the extremely weak coupling to gravity is compensated 
by the additional phase space in the extra dimensions. For this reason gravitational 
phenomena would appear at energies around the $D$-dimensional fundamental scale 
of gravity, $M_D$. The size of the ED in this scenario is allowed by experimental 
data to be almost as large as a millimeter \cite{bib:LEDEXP2}. Since the KK mode 
separation relates to the size as $\Delta m \sim 1/R$, the discrete KK series 
can be approximated by a continuous spectrum. In order to simplify the calculations, 
the extra dimensions are assumed to have the geometry of a $n$-dimensional torus.
In this model of gravity in large extra dimensions, the size of the extra dimensions 
($R$) and the fundamental scale of gravity ($M_D$) are related to the Newton constant 
for $n$ extra dimensions as,
\begin{equation}
 G_N^{-1} = C \cdot R^{n} M_D^{n+2}
\label{eq1}
\end{equation}
where the exact definitions vary slightly between different papers. This work follows 
the parameter definitions used in \cite{bib:LED1} where $C=8\pi$. However, to 
simplify for comparison between papers the different conventions are summarized. 
In the papers \cite{bib:LED1, bib:LED2, bib:LED4} the fundamental scale of gravity 
is referred to as $M_{D}, M$ and $M_{S}$. These are the three most common definitions 
and are related as follows,
\begin{eqnarray}
M^{n+2} = 2 M^{n+2}_D  \\
M^{n+2}_S = 8 \pi^{1-\frac{n}{2}}\Gamma \left ( \frac{n}{2} \right ) M^{n+2}_D 
\end{eqnarray}
The work in \cite{bib:LED4} also defines the size $R$ slightly different compared to 
\cite{bib:LED1, bib:LED2}, $R_S = 2 \pi R_D$. 
From these relations the different expressions in the literature for the integral 
over the very dense KK states reduce to,
\begin{equation}
\sum_k \rightarrow \int dm^2 ~ \frac{\pi^{n/2} \cdot R_D^n \cdot (m^2)^{n/2-1}}{\Gamma(\frac{n}{2})}
\end{equation}
Here $k$ is the graviton momentum component in the extra dimensions and $m^2$ is the 
graviton mass squared.

The unparticle model studied here is described in \cite{bib:UNP2}. In this scenario 
the so-called unparticles originate from a scale invariant sector with a non-trivial 
fixed point. This sector interacts with the SM fields through a connector sector 
with a high mass scale, $M_U$. Renormalization effects in the scale invariant sector 
give rise to dimensional transmutation of the unparticles at an energy scale $\Lambda _U$. 
This transmutation is determined by a scale dimension parameter $d_U$ which has to 
be greater than one due to unitarity arguments, but is allowed to take non-integer 
values. This scenario gives rise to quite unusual phenomena. In terms of real unparticle 
emission it implies that the unparticle would not have a fixed invariant mass squared, 
but instead an invariant mass spectrum. The unparticle would also appear with a 
$d_U$-body final state phase space, which would give rise to an unusual missing energy 
signature. In accordance with \cite{bib:UNP2} only the production of unparticles from 
the allowed spin-1 and spin-2 effective unparticle operators is considered here and 
the unparticles are assumed to be SM singlets. In this model the cross sections are 
determined by the three unparticle parameters, $d_U$, $\Lambda _U$ and $\lambda$. 
Here $\lambda$ is an effective coupling between the unparticle and SM operators which 
is related to the connector scale $M_U$. In the case of spin-2 unparticles two 
different effective operators are allowed which are associated with the independent 
coupling constants $\lambda$ and $\lambda '$. 

As pointed out in \cite{bib:UNP2} the effective theory of spin-2 unparticle emission 
with $\lambda = \lambda '$ is, apart from constant factors, equivalent to graviton 
emission in large extra dimensions. Here the invariant mass spectrum from the dimensional 
transmutation of the unparticles is identical to the dense KK tower of the large extra 
dimensions. For this reason the graviton emission process can be seen as a special 
case of the unparticle model and, as will be described below, both scenarios can be 
covered by the same implementation with only minor changes. 
This has the advantage of allowing the more motivated extra dimension search to be 
extended to cover the unparticle case. To cover both scenarios at once could be 
of great value in the case where future developments discover new and perhaps even 
more interesting analogies between the unparticle picture and gravity in extra 
dimensions. A potential connection between a non-integer $d_U$ value and so-called 
warped extra dimensions has for example been addressed in \cite{bib:UNP3}.

\section{Process Implementation}

The process $\antibar{f} \rightarrow \ZGU $ was implemented\footnote{As a by-product 
also the process $\antibar{f} \rightarrow \gamma + G/U$ was implemented in a separate 
class. This corresponds to the photon limit of the $Z$ process as described later in the 
text.} as a semi-internal process to Pythia 8.108. It contains a 2-to-2 parton level process 
class which can be used by the main Pythia library. The class is implemented with the 
same structure as used for the internal parton level processes and is, therefore, also 
used in the same way as the internal classes inside the Pythia library. No $Z/\gamma^*$ 
interference effects were taken into account and the $Z$ decays isotropically. The 
location and details of the software can be found in the appendix. 

The differential cross section used corresponds to 
\begin{equation}
\frac{d \sigma}{dp^2_Udt} {(\antibar{f} \rightarrow Z + U )} = 
\frac{|{\cal \overline{M}} |^2}{16 \pi \cdot s^2} \cdot \frac{A(d_U)}
     {2 \pi \cdot \Lambda _{U}^{2}} 
     \left ( \frac{p_U^2}{\Lambda _{U}^{2}} \right ) ^{d_U-2} \theta(p^0_U) \theta(p_U^2)
\label{eq:xs}
\end{equation}
where $p^2_U$ is the invariant mass squared. Apart from the matrix element, the first 
term only contains standard 2-to-2 scattering 
phase space factors. The remaining terms contain the phase space and final state mass 
spectrum of the unparticle. Here, $A(d_U)$ is a normalization constant of the $d_U$-body 
phase space. The full spin- and  color-averaged matrix elements given by equations (42) 
and (47) in \cite{bib:UNP2} are used. This corresponds to the most general form of the 
matrix element from the two allowed effective spin-2 unparticle operators. The spin-1 
case only includes the contribution from the vectorial operator. This is sufficient when 
considering unpolarized particle beams, since including the second axial-vectorial operator 
would result in the same formula but with $\lambda \rightarrow \sqrt{\lambda_v^2 + \lambda_a^2}$.

The variable mass spectrum shown in equation (\ref{eq:xs}) gave rise to the main 
difference with respect to the internal processes available in Pythia. It turned out, 
however, to be conveniently implemented by re-weighting an internally produced Breit-Wigner 
distribution. By using the graviton already available in Pythia, the unparticle masses 
are generated according to a Breit-Wigner which is re-weighted to a flat distribution. 
The invariant mass measure of the cross section, $(p_U^2)^{d_U-2}dp_U^2$, then ensures 
that the final events are generated with the correct mass distribution.
In order to achieve a high MC efficiency one has to take care to match the bulk of the 
Breit-Wigner to the bulk of the unparticle distribution. It should be pointed out that 
this only affects the speed of generating events, by throwing events more often in the 
region where the differential cross section is large, and not the final cross section 
or event properties.

The same formula and code is used for producing graviton events. In this case 
$\lambda = \lambda ' = 1$ is used independent of the input values and the $d_U$-body 
phase space factor is changed to,
\begin{equation}
A(d_U) = \frac{16 \pi^2 \sqrt{\pi}}{(2\pi)^{2d_U}}
\frac{\Gamma(d_U + \frac{1}{2})}{\Gamma(d_U - 1)\Gamma(2d_U)} 
\rightarrow S'(n) = \frac{2 \pi \cdot \pi^{n/2}}{\Gamma(\frac{n}{2})}
\label{eq:gc1}
\end{equation}
The remaining input arguments are used in the following way,
\begin{eqnarray}
d_U = \frac{n}{2}+1 \label{eq:gc2} \\ 
\Lambda _U = M_D \label{eq:gc3}
\end{eqnarray}
which implies that in total only two constants have to be changed in order to switch 
between the models.

\begin{figure}[htb]
\begin{center}
 \epsfxsize=7.5cm
 \epsfbox{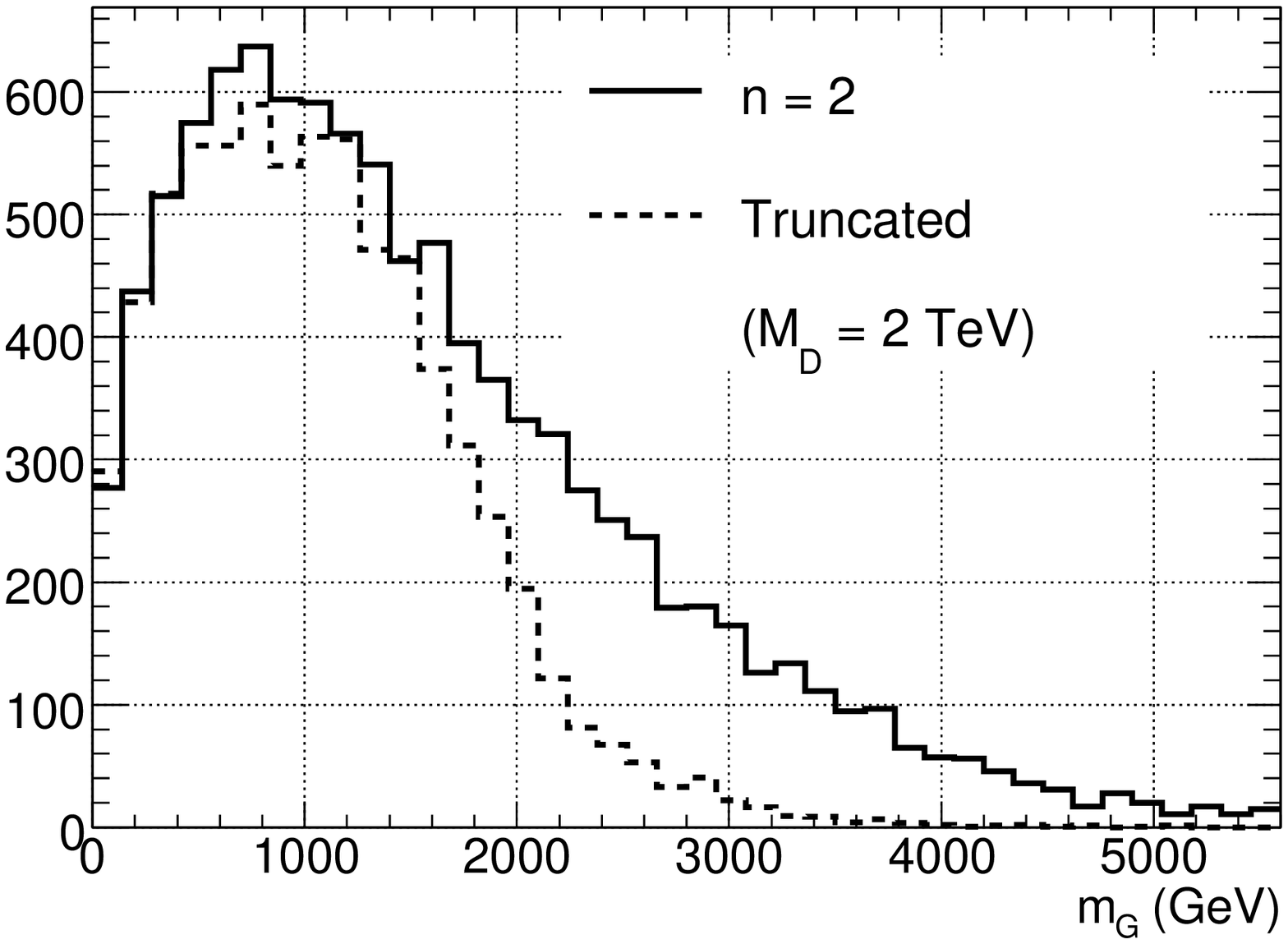}
 \epsfxsize=7.5cm
 \epsfbox{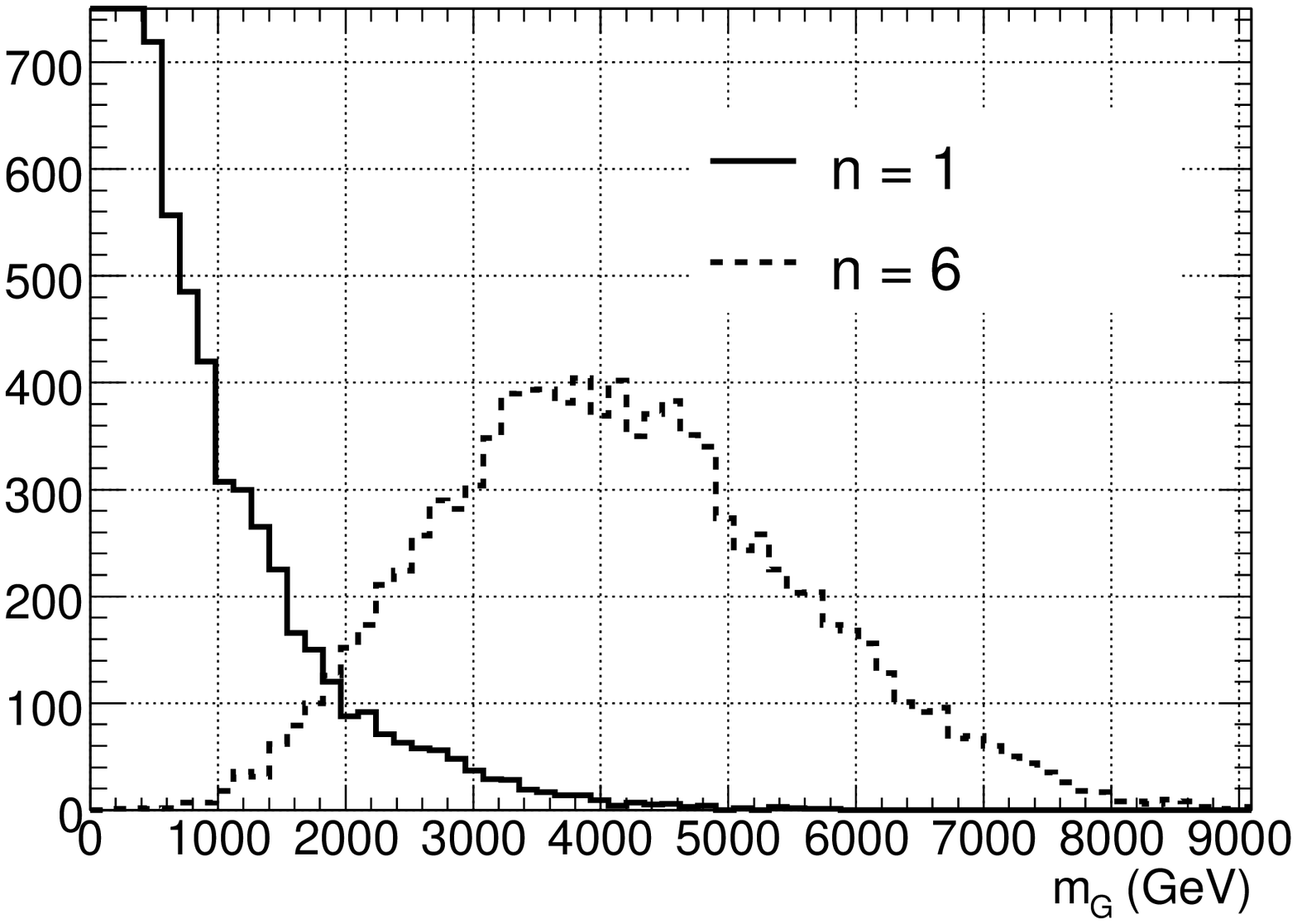}
\end{center}
\caption{\small The graviton mass spectrum for $n = 2$ with and without truncation (left) 
as well as for $n = 1$ and 6 with no truncation (right).}
\label{fig:massspect}
\end{figure}

A truncation option was also implemented in order to test the validity of the effective
theory. In order to trust the perturbative calculations of the effective theory the 
$\hat{s}$ of the process should be smaller than the fundamental mass scale of the theory, 
$\Lambda_U$ or $M_D$. This does not have to be the case at the LHC where the collision 
energy is $14 ~\TeV$ and, for example, the $M_D$ scale at present does not have to be 
larger than about $1~\TeV$. For this reason the truncation functionality suppresses the 
cross section at $\hat{s} > \Lambda_U^2$ by a factor $\Lambda_U^4/\hat{s}^2$. This truncation 
also implies that the mass spectrum is suppressed at large values. The truncation effect 
becomes increasingly significant with large $d_U$ values, since the mass spectrum then 
becomes peaked towards higher values. This is illustrated in figure \ref{fig:massspect}, 
which shows the graviton mass spectrum for $n = 2$ with and without truncation (left) as 
well as for $n = 1$ and 6 with no truncation (right). The truncation in the case $n = 2$ 
is using $M_D = 2 ~ \TeV$. In a similar way the truncation effect becomes increasingly 
important with an increased transverse energy requirement of the event selection in the 
analysis.

The truncation option was primarily used to verify that the difference between the cross 
section obtained with and without truncation is negligible. This implies that the 
uncontrolled $\hat{s}$--region does not contribute significantly. However, in the case of 
a discrepancy, the truncated cross section can still be used as a conservative estimate 
from the region which is under perturbative control.

\section{Validation Cross Checks}

In order to validate the different parts of the implementation, the results were checked 
against several similar processes:
\begin{itemize}
\item{$\ee \rightarrow \gamma + G$ in \cite{bib:LED1, bib:LED2}};
\item{$\pp \rightarrow \gamma + G$ in \cite{bib:LED1}};
\item{$\ee \rightarrow \ZG $ in \cite{bib:LED3, bib:LED4}};
\item{$\ee \rightarrow \gamma + U$ in \cite{bib:UNP2}}.
\end{itemize}

Together these should provide sufficient cross checks to validate all parts of the 
implemented process. Most parts are covered by reproducing the results in 
\cite{bib:LED1, bib:LED2} from the first two processes. The third processes verifies the 
$Z$ specific parts of the spin-2 matrix element. The fourth process verifies that the 
unparticle phase space factors and spin-1 matrix element are correct. The only part which 
is not directly tested is the matrix element in the case of a spin-2 unparticle with 
$\lambda \neq \lambda '$. On the other hand, when setting $\lambda = \lambda '$ a specific 
cancellation is required in order to reproduce the \ZG\ matrix element and therefore this 
is considered to be a good general test.

The same code was also used to produce the photon results. The photon limit of the \ZGU\ 
process was obtained by making the following changes,
\begin{eqnarray}
m_Z \rightarrow 0 \\
\frac{g_v^2 + g_a^2}{4} \rightarrow Q^2 \\
\frac{g}{\cos{\theta_W}} \rightarrow e 
\end{eqnarray}
where the following coupling conventions were used, $g_v = -\frac{1}{2}+2\sin^2{\theta _W}$ 
and $g_a = -\frac{1}{2}$. The fact that the photon limit of the $Z$ matrix element is 
checked is also a valuable cross check of the implementation.

\section{\ZGU\ Production at the LHC}

The potential of using \ZGU\ events to confirm a signal observed in the main channels was 
also studied. These events would give rise to di-leptons and large missing transverse energy. 
The most favorable event topology from an experimental point of view is when the leptons are 
either an electron or muon pair. The SM process with the most similar experimental signature 
is $ZZ$ production, where one $Z$ decays into electrons or muons and the other into neutrinos, 
\ZZdecay . For this reason it is interesting to investigate the possible \ZGU\ cross section 
relative to this SM \ZZ\ process. 

The ATLAS \ZZdecay\ analysis presented in \cite{bib:ZZATLAS} will be used as a reference 
through this study. This analysis predicts an overall inclusive \ZZ\ cross section of 
$\sigma _{ZZ} = 14.8~pb$, which was obtained by the {\tt MC@NLO} generator and using the 
{\tt CTEQ6M} parton distribution functions (PDFs). The corresponding cross section obtained 
by Pythia 8 was $11.4~pb$ using the {\tt CTEQ5L} PDFs. This appears consistent for a comparison 
between leading order (LO) and next-to-leading order (NLO) results.

The \ZZ\ events generated by Pythia were then passed through a similar event selection to the 
one used in \cite{bib:ZZATLAS}. This corresponds to the following cuts:
\begin{itemize}
\item[1)]{Opposite charged electrons or muons;}
\item[2)]{$p^{\ell}_T > 20 ~GeV$;}
\item[3)]{$|\eta^{\ell}| < 2.5$;}
\item[4)]{$|M_{\ell \ell} - 91.2| < 10 ~GeV$;}
\item[5)]{$p_T(Z) > 100 ~GeV$.}
\end{itemize}
The first three requirements ensure that there are two high-$p_T$ electrons or muons inside 
the central part of the experiment. The remaining two cuts imply that the leptons are consistent 
with the decay of an on-shell $Z$ with high $p_T$. After applying this selection to the 
generator level \ZZ\ events from Pythia, a selected cross section of $14 ~fb$ was obtained. 
This value also appears reasonable compared to the $10.2 \pm 0.2 ~fb$ obtained in the full 
ATLAS analysis, when considering that they are calculated at LO and NLO respectively, and that 
the ATLAS analysis includes the full ATLAS detector simulation and reconstruction chain.

Since the existing limits on $M_D$ exclude values between approximately $0.5$ and $1.5 ~\TeV$ 
the following parameter values were used as benchmark points for the \ZG\ study: 
\begin{itemize}
\item{$n = 1,~2,~3,~6$;}
\item{$M_D = 2, ~2.5, ~3 ~\TeV$.}
\end{itemize} 
Events generated with, $n=2$, $M_D = 2 ~\TeV $ and using the truncation option were passed 
through the \ZZ\ selection described above and a selected cross section of $2.6 ~fb$ was 
obtained. This shows that for this particular point the \ZG\ signal would amount to about 
20\% of the expected SM \ZZdecay\ signal. 

\begin{figure}[htb]
\begin{center}
 \epsfxsize=16cm
\epsfbox{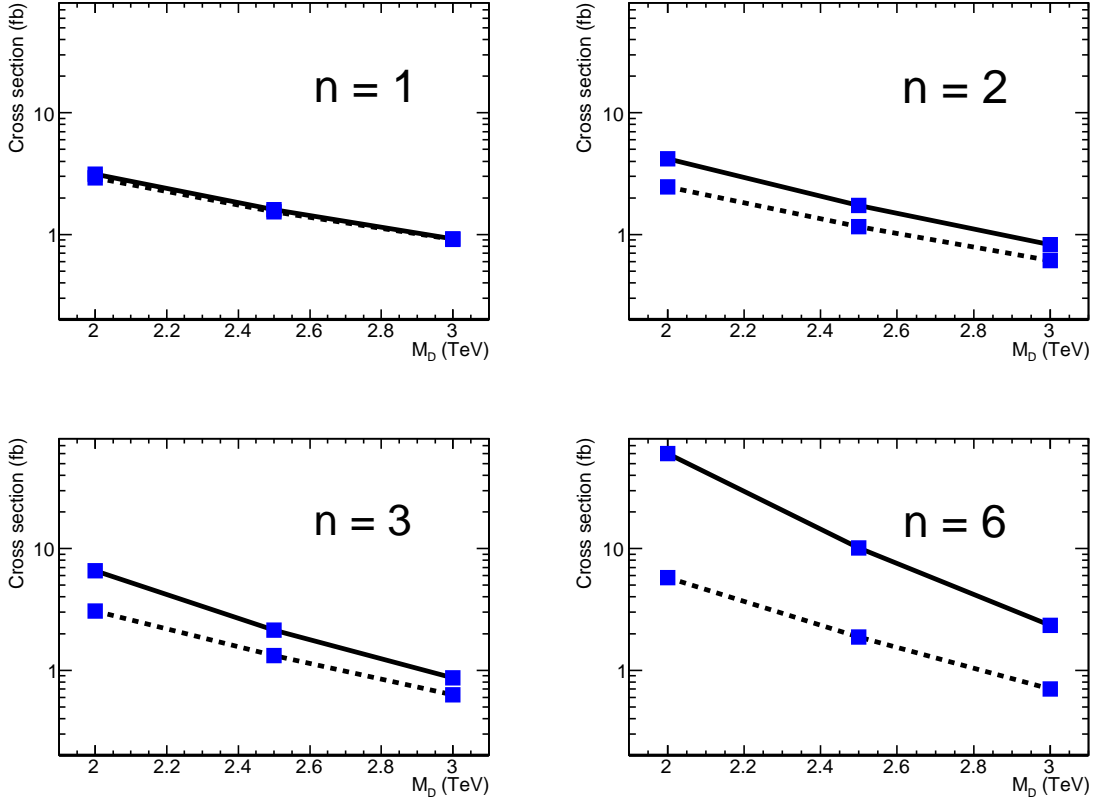}
\end{center}
\caption{\small Selected \ZG\ cross section as a function of $M_D$. The results are shown 
with (dashed line) and without (solid line) truncation.}
\label{fig:zgxs}
\end{figure}

Figure \ref{fig:zgxs} shows the selected cross section for the different benchmark points 
and the results are shown with and without truncation. One can clearly see that the cross 
section increases with a larger number of extra dimensions, but at the same time an increasing 
fraction moves into the non-perturbative $\hat{s}$ region. For this reason the effective 
theory is not trustworthy at large $n$. At small $n$, on the other hand, the effective 
theory is fairly valid and the \ZG\ channel could be used to cross check a signal observed 
in the more conventional channels. This will, however, require relatively large amounts of 
data. Assuming for example $\sigma_b = 20 ~fb$ of selected background\footnote{This background 
value is assumed since the ATLAS study suggests about 50\% more selected \ZZ\ candidates 
than real \ZZ\ events.} in total and a signal 
of $\sigma_s = 2.6 ~fb$, a data sample in the order of,  
\begin{equation}
{\cal L} = \frac{25 \cdot \sigma_b}{\sigma_s^2} \sim 75 ~ fb^{-1}  
\end{equation}
would be necessary for a $5\sigma$ discovery.

\begin{figure}[htb]
\begin{center}
 \epsfxsize=7.9cm
 \epsfbox{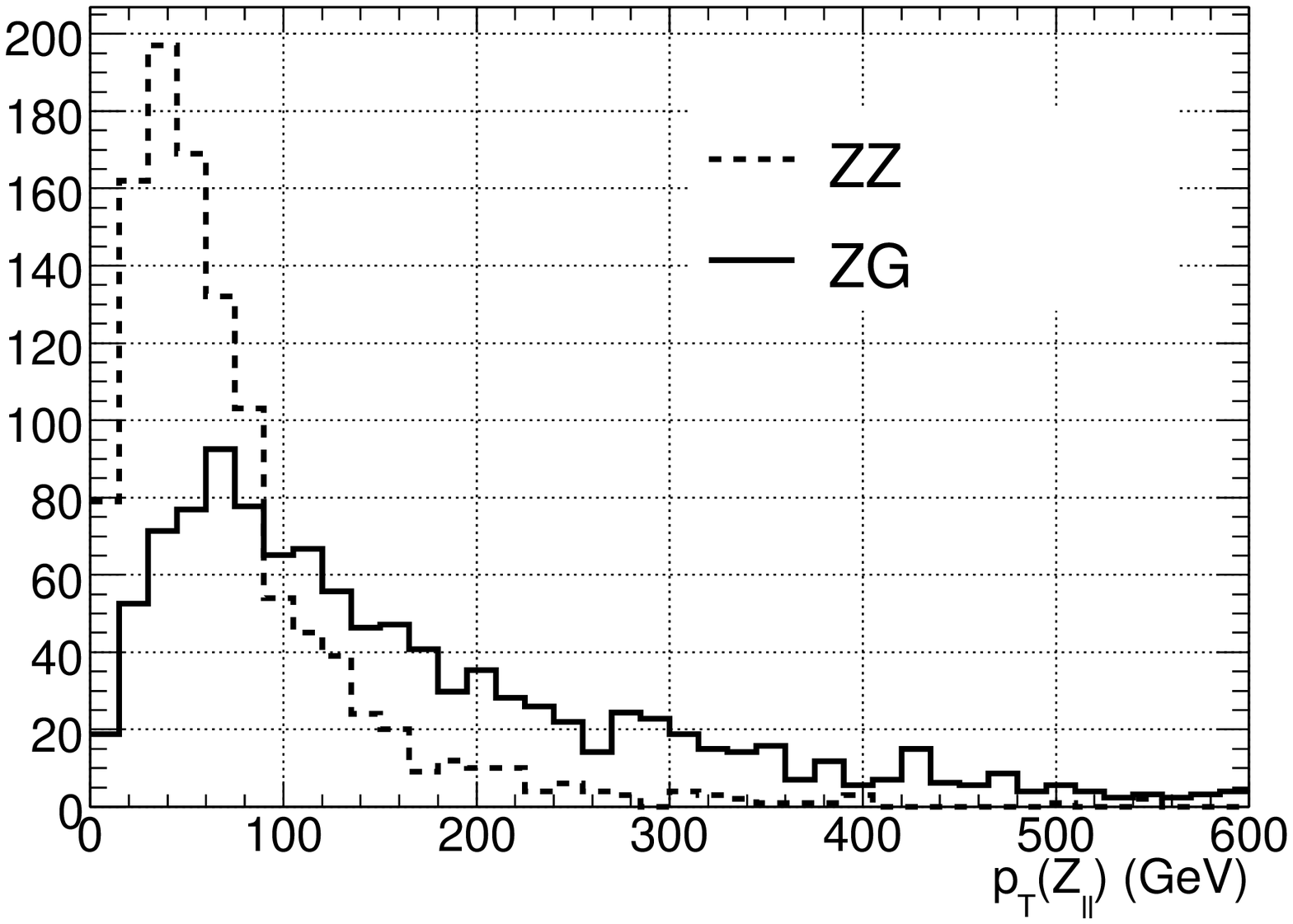}
 \epsfxsize=7.9cm
 \epsfbox{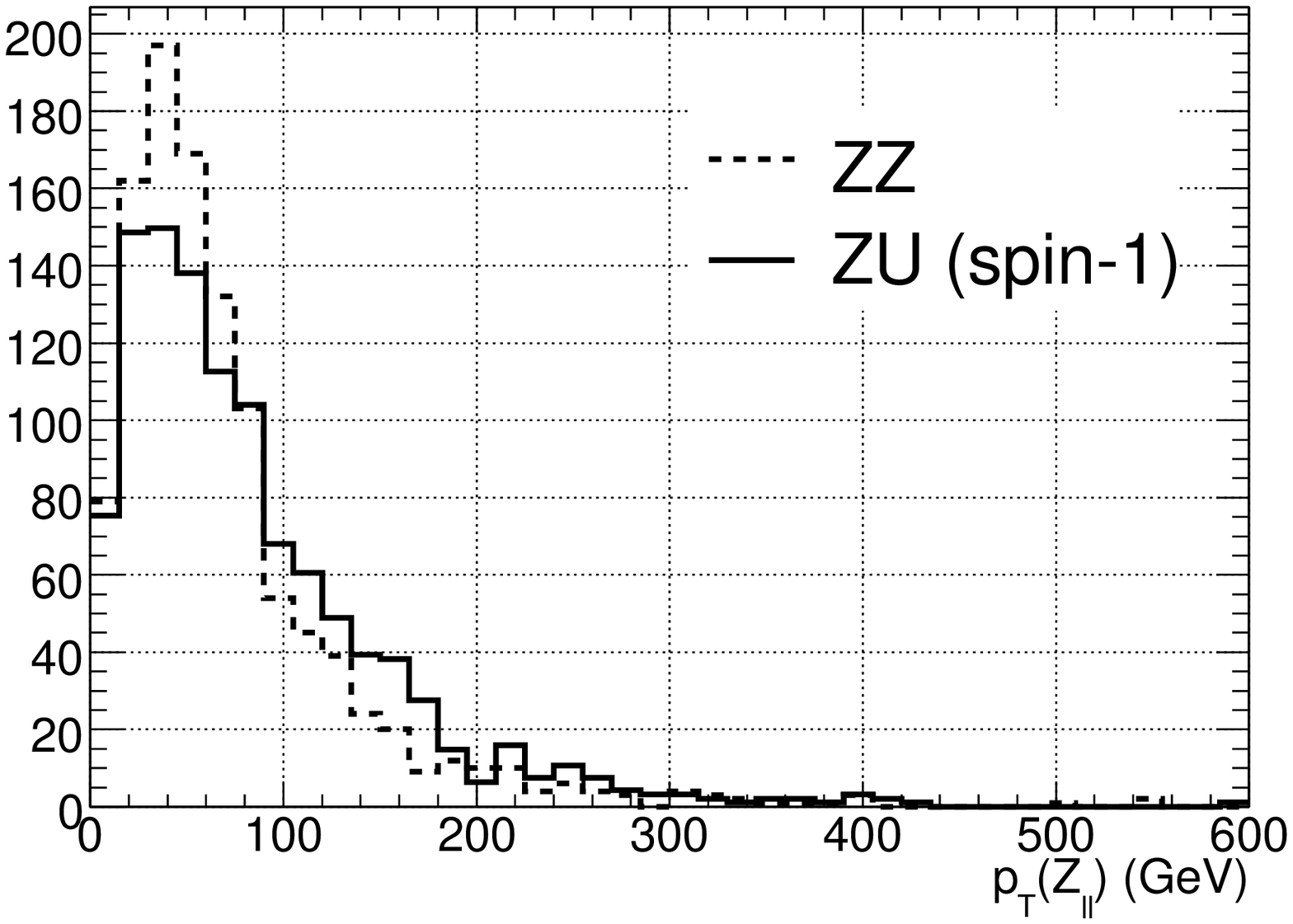}
\end{center}
\caption{\small Comparison between the $p_T(Z)$ spectra from the SM \ZZdecay\ and the 
\ZGU\ processes including truncation. The left plot corresponds to a \ZG\ signal with 
$n=2$ and $M_D = 2 ~ \TeV $. The right plot corresponds to a spin-1 \ZU\ signal with 
$d_U = 2$ and $\Lambda_U = 2 ~ \TeV $. All distributions are based on events with two 
electrons or muons that pass the cuts 1 to 3 and have the same arbitrary normalization.}
\label{fig:met}
\end{figure}

The event selection could be improved to increase the sensitivity of the graviton 
signal. The left plot of figure \ref{fig:met} shows the $p_T(Z)$ distributions of the SM 
\ZZdecay\ process and the \ZG\ process with $n=2$ and $M_D = 2 ~ \TeV $. Here a different 
shape can be seen and this could, for example, be used to optimize the selection. It was 
also observed that the $p_T(Z)$ spectrum becomes harder with an increasing value of $n$.
A detailed study of the selection should, however, include proper detector simulation and 
event reconstruction. This will not be covered here, but would be a natural continuation 
of this work.
 
The same selection was used to investigate the case of spin-1 unparticle emission.
Here the \ZU\ signal was generated using the parameter points $d_U = 1.4, ~1.6$ and 
$2.0$ in accordance with \cite{bib:UNP2} and $\lambda =1$ was kept fixed.
Despite the fact that the unparticle mass spectrum shown in equation (\ref{eq:xs}) is 
the same as in the spin-2 case, the spin-1 matrix element implies a much softer $m_U$ 
as well as $p_T(Z)$ distribution. The $p_T(Z)$ distribution is illustrated in the right 
plot of figure \ref{fig:met}, which shows spin-1 unparticles with $d_U = 2$ compared to 
the SM \ZZ\ distribution. For this reason the detection efficiency is significantly lower 
than in the spin-2 case and, even with large amounts of data, the prospects seem poor for 
reaching significantly beyond the LEPII limits \cite{bib:UNP2} using the $Z+U$ channel at 
the LHC.

The corresponding results for the spin-2 unparticle scenario with $\lambda = \lambda '$ 
can be obtained simply by scaling the graviton results according to, 
\begin{equation}
\frac{\sigma_U}{\sigma_G} = \frac{A(d_U)}{S'(n)} \cdot \lambda ^2
\end{equation}
For this reason, the study of spin-2 unparticles was focused on deviating effects when 
$\lambda \neq \lambda '$. 
\begin{figure}[htb]
\begin{center}
 \epsfxsize=7.9cm
 \epsfbox{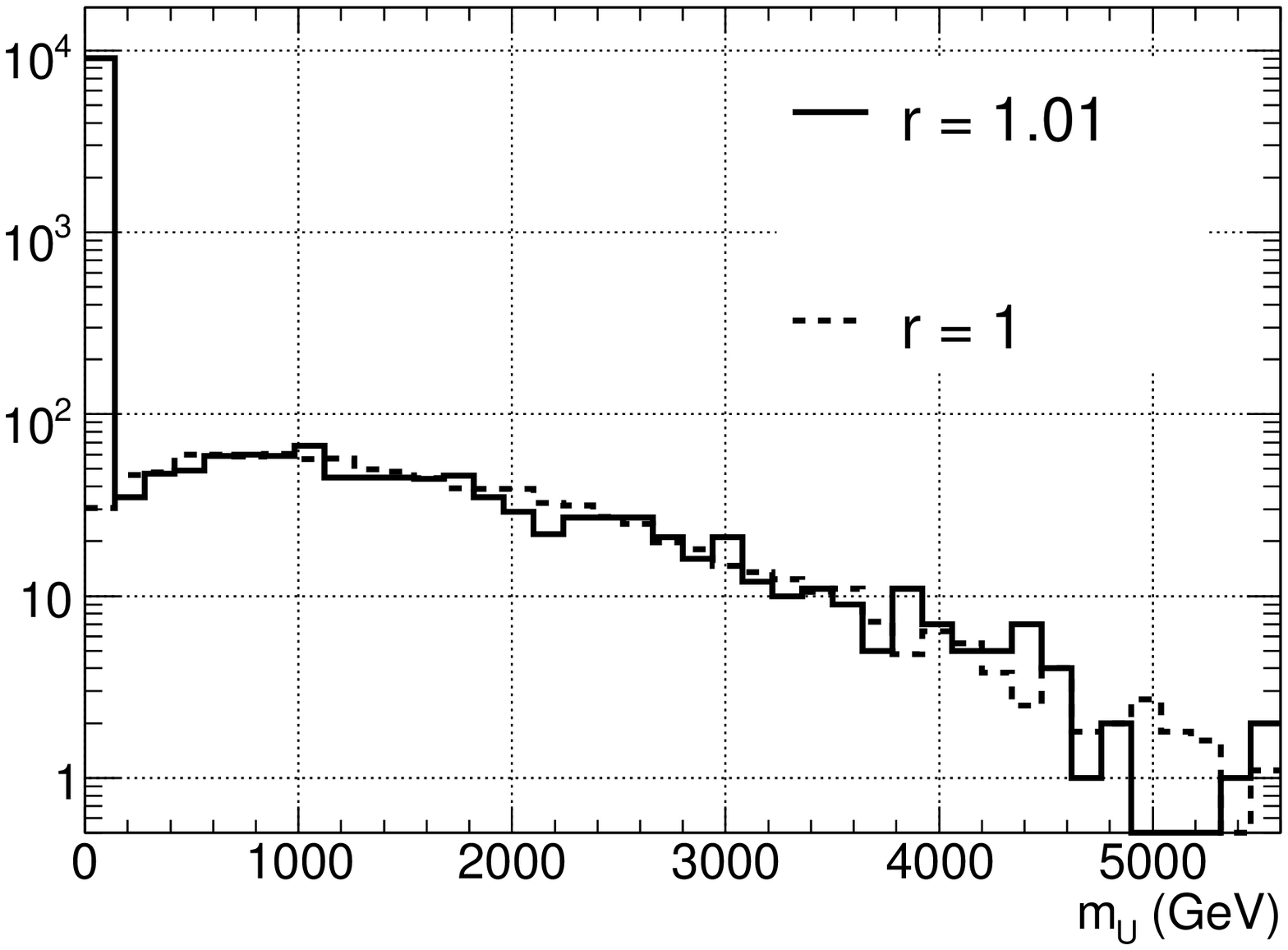}
 \epsfxsize=7.9cm
 \epsfbox{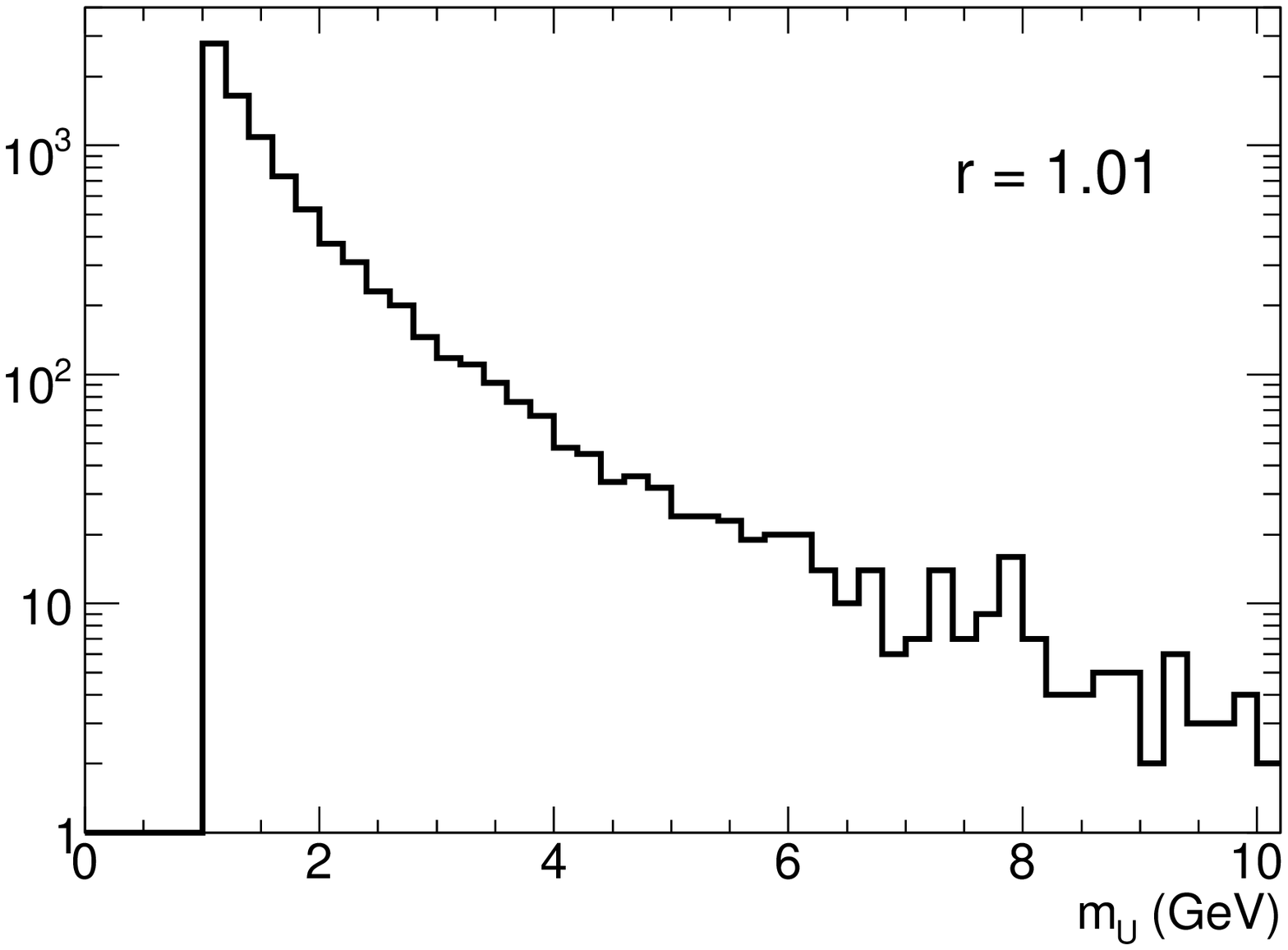}
\end{center}
\caption{\small The (spin-2) unparticle mass spectra for $d_U = 2$ with $r = 1$ and $r = 1.01$. 
The histograms are arbitrarily normalized. The distributions in the left plot, however, have the 
correct ratio with respect to the total cross sections. The cut at 1 \GeV\ in the right plot 
is a free parameter of the MC implementation.}
\label{fig:u2}
\end{figure}
Figure \ref{fig:u2} shows the $m_U$ distributions of events generated with $d_U = 2$ and 
$r [ = \lambda' / \lambda ] = 1.01$ as well as $r = 1$. In the left plot it can be seen 
that even for this small deviation from $\lambda = \lambda '$, a large spike develops in 
the first bin. This spike is resolved in the right plot and indicates an infra-red (IR) 
divergence of the cross section in the case of $\lambda \ne \lambda '$. This is explained 
to some extent by the fact that the terms in the matrix element that cancel when $\lambda = \lambda '$ 
are all proportional to powers of $1/m^2_U$. For this reason also the total inclusive 
cross section increases rapidly when the couplings are not equal. 
Unlike the spin-1 case, the events in the soft $m_U$ tail have a significantly harder 
$p_T(Z)$ spectrum than the SM \ZZ\ process. This indicates that the \ZU\ channel could also 
be a useful complement when investigating scenarios with $\lambda \ne \lambda '$.

\section{Conclusions}

Models with extra dimensions have attracted much interest as candidates for potential 
physics beyond the SM. One of the most popular phenomena is real graviton emission and 
it has been pointed out earlier that graviton emission can be seen as a special case of 
the so-called unparticle emission process. This allows the same MC implementation to 
cover both scenarios and the parton level process $\antibar{f} \rightarrow Z/\gamma + G/U$ 
has been implemented in Pythia 8.
In addition, \ZGU\ production at the LHC has been studied as a complement to the more 
conventional channels. This process was studied at generator level relative to the 
similar SM process \ZZdecay . In order to get a handle on the experimental effects, a 
detailed analysis of the ATLAS \ZZdecay\ cross section measurement has been used as a 
reference. The study indicates that \ZG\ production could be used for 
small values of $n$ to confirm a signal observed in the main channels. It will, however, 
require a relatively large amount of data. Spin-1 unparticle production at the LHC, on the 
other hand, seems more difficult to observe using the \ZU\ channel.
Spin-2 unparticle events have the same characteristics as the graviton events if 
$\lambda = \lambda '$. When these couplings are different an IR divergence is observed 
in the $m_U$ spectrum. This increases the total inclusive cross section and the \ZU\ 
channel could provide a useful complement also in this scenario.

\section{Acknowledgment}

The author wants to express special thanks to Kingman Cheung (Hsinchu) and Gian Giudice 
(CERN) for helpful discussions. Also Torbj$\ddot{\mbox{o}}$rn Sj$\ddot{\mbox{o}}$strand 
(Lund U.) is greatly acknowledged for his help with various Pythia related matters. This 
work was funded in the UK by STFC.

\clearpage
\appendix
\section*{Appendix - Implementation in Pythia 8}

The processes $\antibar{f} \rightarrow \ZGU $ and $\antibar{f} \rightarrow \gamma + G/U $ 
are implemented in two semi-internal classes called {\tt Sigma2ffbar2UZ} and {\tt Sigma2ffbar2Ug}. 
The classes include a number of standard Pythia methods which are called at different 
points during the event generation. For this reason they can be used by the main Pythia 8 
library in the same way as for example the {\tt Sigma2bg2Hb} process in the {\tt main25.cc} 
example of Pythia version 8.108.

The constructors takes 7 arguments in order to instantiate an object,
\begin{itemize}
\item{{\tt int Spin}, unparticle spin (1 or 2);}
\item{{\tt bool Trunc}, set {\tt true} in order to truncate the contributing $\hat{s}$ values;}
\item{{\tt bool Graviton}, set {\tt true} to use graviton specific settings according to eqs. 
\ref{eq:gc1}, \ref{eq:gc2} and \ref{eq:gc3};}
\item{{\tt double} $d_U$, scale dimension parameter;}
\item{{\tt double} $\Lambda_U$, unparticle renormalization scale;}
\item{{\tt double} $\lambda$, unparticle coupling to SM; }
\item{{\tt double Ratio}, the ratio $\lambda ' / \lambda$ of the spin-2 matrix element \cite{bib:UNP2}.}
\end{itemize}
If {\tt Graviton = true}, the $\lambda$ and {\tt Ratio} values are overridden and set 
equal to one.
 
The graviton particle code 39 was used both in the case of graviton and unparticle 
emission. The graviton available in Pythia has a Breit-Wigner distributed mass and 
the underlying Breit-Wigner shape can be adjusted in the main program by the following 
parameters,
\begin{verbatim}
pythia.readString("39:m0 = 50.");
pythia.readString("39:mWidth = 500.");
pythia.readString("39:mMin = 1.");
pythia.readString("39:mMax = 13910.");
\end{verbatim}
The Breit-Wigner was re-weighted to a flat distribution by using the {\tt runBW3} 
weight in the {\tt sigmaHat()} method. By doing this the differential unparticle 
cross section in equation \ref{eq:xs} ensures a proper mass distribution of the 
generated final events. \\

{\bf \noindent Program Summary} \\
{\it
Code Location: sask.home.cern.ch/sask/gravunp100.tgz \\
Tested with: PYTHIA version 8.108 \\

\noindent The .tar file contains: \\
README \\
main\_gravunp.cc \\
Sigma2ffbar2UZ.h \\
Sigma2ffbar2Ug.h \\
Sigma2ffbar2UX.cc \\
}

\clearpage


\end{document}